\documentclass[twoside,letterpaper]{IEEEtran}
\usepackage{graphicx}
\usepackage{subfigure}
\usepackage{booktabs}
\usepackage{paralist}
\usepackage{algorithm} 
\usepackage{algpseudocode}
\usepackage{color}
\usepackage{stfloats}
\usepackage{url}

\renewcommand{\baselinestretch}{1.045}

\begin{document}

\title{Trilateral Large-Scale OSN Account Linkability Study}
\author{Alice Tweeter \and Bob Yelper \and Eve Flickerer}
\maketitle
\begin{abstract}
In the last decade, Online Social Networks (OSNs) have taken the world by storm.
They range from superficial to professional, from focused to general-purpose, and, 
from free-form to highly structured. Numerous people have multiple accounts within
the same OSN and even more people have an account on more than one OSN. 
Since all OSNs involve some amount of user input, often in written form, it is natural to consider
whether multiple incarnations of the same person in various OSNs can be 
effectively correlated or linked. One intuitive means of linking accounts 
is by using stylometric analysis.

This paper reports on (what we believe to be) the first trilateral large-scale
stylometric OSN linkability study. Its outcome has important implications for 
OSN privacy. The study is trilateral since it involves three OSNs with very 
different missions: (1) Yelp, known primarily for its user-contributed reviews of 
various venues, e.g, dining and entertainment, (2) Twitter, popular for its pithy
general-purpose micro-blogging style, and (3) Flickr, used exclusively for posting
and labeling (describing) photographs. As our somewhat surprising results indicate, 
stylometric linkability of accounts across these heterogeneous OSNs is both viable
and quite effective. The main take-away of this work is that, despite OSN 
heterogeneity, it is very challenging for one person to maintain privacy 
across multiple active accounts on different OSNs.

\end{abstract}

\section{Introduction}
\label{sec:introduction}
Online Social Networks (OSNs) have been rapidly gaining worldwide popularity for almost two decades. 
The OSN paradigm evolved from pre-web BBSs (Bulletin Board Systems) and Usenet discussion
groups, through AOL\cite{aol} and Yahoo, to enormous and global modern OSNs.
One of them, Twitter, has already exceeded $200,000,000$ accounts \cite{twitterusers}. 
In addition to gaining users, OSNs have permeated into many spheres of everyday life.
One of many possible ways to classify OSNs is by their primary {\em mission}:
\begin{compactitem}
  \item {\bf Generic OSNs}, such as Facebook, VK, Google+ and LinkedIn, where users establish and maintain 
  connections while sharing any type of content, of almost any size.
  \item {\bf Microblogging OSNs}, such as Twitter and Tumblr, that let users share short,  
  frequent and (ostensibly) news-worthy missives. 
  \item {\bf Media-specific OSNs}, such as Instagram and Flickr, where users mainly share content of a certain 
  media type, such as photos or videos. However, even in these OSNs, users provide textual labels and descriptions 
  for shared media content.
  \item {\bf Review OSNs}, such as Yelp, TridAdvisor and Amazon, where users offer of products and services, 
  e.g. restaurants, hotels, airlines, music, books, etc. These tend to be hybrid sites, that include some social 
  networking functionality, beyond user-provided reviews. Users are evaluated by their reputations
  and there are typically no size restrictions on reviews.
\end{compactitem}
Despite their indisputable popularity, OSNs prompt some privacy concerns.\footnote{This is despite the fact
that the entire notion of ``OSN Privacy'' might seem inherently contradictory.} With growing revenue 
on targeted ads, many OSNs are motivated to increase and broaden user profiling and, in the
process, accumulate large amounts of Personally Identifiable Information (PII). Disclosure of this PII, whether 
accidental or intentional, can have unpleasant and even disastrous consequences for some OSN users. 
Many OSNs acknowledge this concern  offering adjustable settings for desired privacy levels.

\subsection{Motivation}
\label{subsec:motivation}
Meanwhile, a large number of people have accounts on multiple OSNs, especially, OSNs of different types. 
For example, it is common for someone in his/her 20-s to have a Twitter, Instagram and Facebook
accounts. However, privacy {\bf across} OSNs is not yet sufficiently explored. Many users
naturally expect that their accumulated contributions (content) and behavior in one OSN account are
confined to that OSN. It would be clearly detrimental to one's privacy if correlating or linking accounts 
of the same person across OSNs were possible. 
 
\begin{figure}[t]
  \fbox{\centering
  \includegraphics[width=0.47\textwidth]{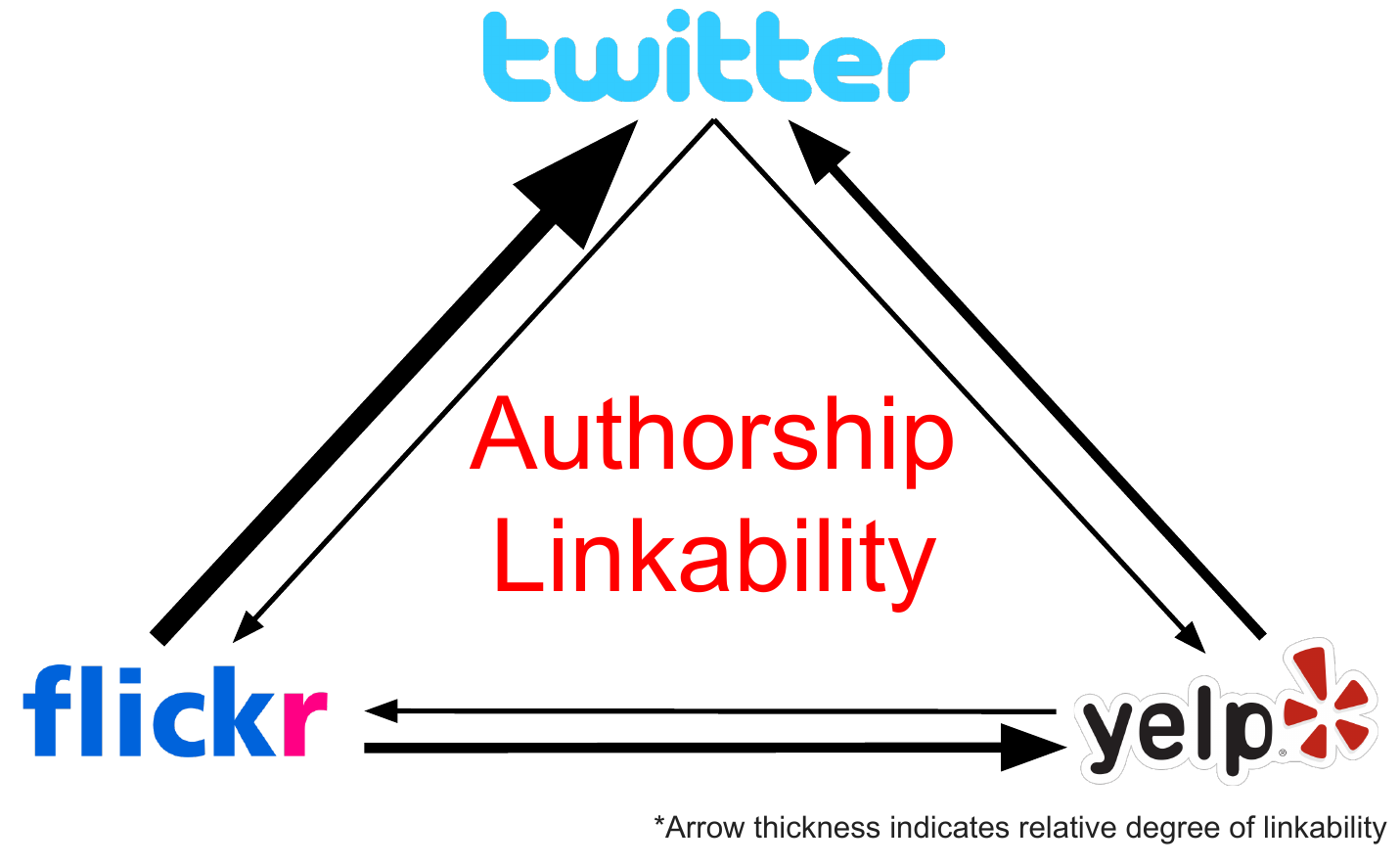}}
  \caption{Summary of the trilateral OSN account linkability study.}
  {\label{fig:trilateral-linkability}}
\end{figure}

In this paper, we explore linkability of user accounts across OSNs of different types. That is, given a user holding 
accounts on two OSNs, we investigate the efficacy and efforts needed to correctly link these accounts.  While this 
problem has been studied in \cite{goga2013exploiting}, prior results are very limited with respect to linkage
accuracy and large numbers of accounts. The goal of this work is to develop cross-OSN linkage models that are 
highly accurate and scalable. To this end, we apply \textit{Stylometry} -- the study of one's writing -- in a novel framework, 
that yields very encouraging results. Our linkability study is performed over three popular OSNs: Twitter, Yelp and Flickr. 
These OSNs are heterogeneous, i.e.,  each has a very distinct primary mission. Thus, the problem of accurately 
linking users accounts is quite challenging. Figure \ref{fig:trilateral-linkability} captures the 
OSN pairs we study for linkability purposes. 

Although accurate and scalable linking techniques are detrimental to user privacy, they can also be useful in 
forensics, e.g., to trace various miscreants. As is well-known, OSNs have become a favorite global 
media outlet for both criminals and terrorists to recruit and promote ideology. Both sides of linkability arguments 
are equally important. However, we believe that it is important to know potential privacy consequences of 
participating in multiple OSNs, since, as mentioned above, many (perhaps na\"ively) expect some confinement
or compartmentalization of each OSN account.

\subsection{Contributions} 
\label{subsec:contributions}
Our anticipated contributions are as follows:
\begin{compactitem}
  \item \textbf{High Accuracy.}
   We develop stylometric-based linkability models that are substantially more accurate than those in previous work, e.g,      
   \cite{goga2013exploiting}. 
   \item \textbf{Scalability.}  Popular OSNs have enormous numbers of users. Thus, scalability of 
   linkability models is essential.  Unlike previous work, our models easily scale from $100$ to $100,000$ users. 
   \item \textbf{Public Data.} Proposed linkability models perform very well with respect to accuracy and scalability 
   even though we assume that the adversary only has access to publicly available 
   textual data from OSNs.\footnote{We believe that users who pursue privacy would disable all OSN meta-data 
   information, such as geo-location
   -- a feature that was essential for linkability accuracy in \cite{goga2013exploiting}.
   Moreover, private messages will not be available to outside world,
   which was used in \cite{afroz2014doppelganger}.}
   Therefore, achieving high accuracy armed only with publicly available data, provides a lower bound on how much 
   the adversary can achieve and serves as an indicator of the severity of the privacy problem.
\end{compactitem}

\subsection{Organization}
\label{subsec:organization}
The rest of the paper is organized as follows:
Section \ref{sec:related} summarizes related work in authorship attribution and linkability. 
Then, Section \ref{sec:background} provides background information on OSNs used in our study.
Problem settings are presented in Section \ref{sec:problem-setting}, followed by 
Section \ref{sec:dataset} which describes the massive dataset used as input for the study.
Section \ref{sec:preliminaries} describes some preliminaries of the experiment framework.
Next, experimental results are presented in Section \ref{sec:results}.
Potential issues and questions stemming from the study results are discussed in 
Section \ref{sec:discussion}. Finally, Section \ref{sec:conclusion} summarizes the paper.
\section{Related Work}
\label{sec:related}
\noindent{\bf Author Attribution.}
There has been a lot of research in the field of author attribution. Abbasi, et al. 
\cite{abbasi2008writeprints} proposed a technique based on a new unsupervised
learning method, referred to as Writeprints. It uses Karhunen-Loeve transforms 
along with a rich set of features to identify authors, achieving accuracy of 91\% 
in finding the author of an anonymous message from a set of 100 candidate authors. 
A study called Herbert West -- Deanonymizer, was conducted to investigate the
possibility of de-anonymizing peer reviews of academic papers 
\cite{herbert-deanonymizer}. A high percentage -- around 90\% -- of reviews 
were correctly de-anonymized from a set of 23 reviewers using Naieve Bayes 
Classifier. Another recent effort studied author identification of the Internet blogs 
on a relatively large-scale, with $100,000$ authors \cite{internet-scale}. In certain 
cases, de-anonymization accuracy of 80\% was achieved and anonymous texts 
were linked cross different platforms. Mishari, et al. \cite{mish-linkability} studied linkability of community-based reviews in Yelp, based on a set of about $2,000$ reviewers and almost all reviews were correctly de-anonymized. Even though
a simple feature set was used (e.g., unigrams and bigrams) with Naive 
Bayesian classifier, high linkability accuracy was achieved.  Stamatatos 
\cite{author-survey} extensively surveyed the area of author attribution  and we 
refer to it for a good overview of the topic.  

\noindent{\bf Cross-Linking Accounts.}
The study most relevant to this paper was conducted by Goga, at al.  \cite{goga2013exploiting}. It cross-linked accounts between different 
OSNs, the same three that are used in this paper: Twitter, Yelp and Flicker. 
Features that included locations, timestamps and text were used, with the
help of the cosine distance function, to link accounts operated by the same 
user across OSNs. While the settings is similar to ours \footnote{As acknowledged 
in Section \ref{sec:dataset}, we borrowed our dataset from this study.}, we substantially 
improve linkability results. Unlike \cite{goga2013exploiting}, we rely only on text-based 
features and leverage them to improve scalability (larger set of accounts) 
and linkability results.
Moreover, we report on correlations in between all OSN pairs,
whereas \cite{goga2013exploiting} only discusses correlating Yelp and Flickr to Twitter.

Similarly, Afroz, et al. \cite{afroz2014doppelganger} successfully explored cross-linking 
multiple accounts belonging to the same user within the same forum or blog-based site.
This is a step forward since, in prior studies, linking was based on artificially 
created accounts of the same user. Accuracy between 85\% and 90\% was 
achieved, while maintaining high recall values. The study used an algorithm called 
{\bf Doppelg{\"a}nger Finder}, where two accounts: $account_A$ 
and $account_B$ were claimed to belong to the same user if combined 
probability of attributing $account_A$ to $account_B$ and vice versa 
exceeded a specific threshold. The probability of attributing $account_A$ to 
$account_B$  was computed based on a model  trained on all accounts except $account_A$ and vice versa. Probabilities are combined by averaging, multiplying 
or square-averaging. Lexical, domain and syntactic features were used along with
Principal Component Analysis to reduce the feature set size.  

A large-scale ($10,000$) author attribution study was recently conducted to link Twitter accounts 
based on very simple lexical features -- unigrams and bigrams -- and Naive Bayesian 
classifier \cite{mish-wpes-14}.  High linkability results -- nearly 100\% -- have been 
achieved. Also, results were verified based on {\em ground truth} -- actual 
Twitter accounts belong to the same user.

Other related work explored account linkability in online services based on entropy of user-names \cite{dperitoPETS11}. 
In \cite{footprint}, account properties with a simple set of heuristics were used to cross-link users.
And finally, Iofciu, et al. explored tags to identify users across Delicious, StumbleUpon and Flickr \cite{iofciu2011identifying}.

\noindent{\bf De-anonymization in User Preference Databases.}
More distantly related is the body of research that addressed de-anonymization 
of contributors to user preference databases. One seminal work studied 
de-anonymization of Netflix database users who rated movies 
\cite{deanonymize-netflix}. 
It proposed a model for privacy breaches, based on an external knowledge base,  
and demonstrated an actual attack on the Netflix dataset. A closely related effort 
proposed techniques for cross-linking user accounts between movie rating 
database and public forums \cite{you-what-u-say}. 

\section{OSN Background}
\label{sec:background}
In this section, we overview three OSNs used in our study.

\noindent{\bf Yelp} is a community-based review site \cite{yelp} 
where users -- who must have accounts -- offer reviews of various products 
and services. Access to reviews is not restricted, i.e., anyone can read Yelp
reviews, with or without an account. Typical reviewed industry categories 
include: restaurants, automotive, medical, hospitality and entertainment. 
At least in North America, Yelp is very popular: the number of reviews 
exceeds $70,000,000$ and the number of yearly visitors is about $142,000,000$ 
\cite{yelpsabout}. Yelp is considered to be an OSN since it also allows
its users to connect to, and interact with, other Yelp users. 
Yelp has a reward system for reviewers based on the quantity and quality
(popularity and ratings) of their contributions. Not surprisingly, this helps 
increase the number of avid or prolific reviewers \cite{yelpeliete}. 

\noindent{\bf Twitter} is a microblogging OSN \cite{twitter} where 
registered users (known as tweeters) post short messages (called 
tweets).\footnote{Technically, one can be a Twitter user but not a tweeter,
e.g., someone might create an account only to follow others' tweets, but not tweet.}
Some tweeters make their tweets public, meaning that anyone can
read them regardless of having a Twitter account. Meanwhile, others 
restrict access to their tweets to so-called followers -- Twitter users who 
have explicitly requested, and have been
granted, access to one's tweets. One of Twitter's most distinctive features
is the $140$-character size limit for tweets. Twitter is currently one of the most 
popular and diverse OSNs, having attracted many avid tweeters among
politicians, journalists, athletes and various celebrities. Furthermore, all kinds
of groups, societies and organizations (both in public and private sectors) 
have strong Twitter presence. The number of Twitter accounts exceeds $200,000,000$ \cite{twitterusers}.   

\noindent{\bf Flickr} is a focused OSN and a cloud storage provider, 
specializing in sharing multimedia content, i.e., photographs and videos \cite{flickr}. 
Flickr users can annotate their multimedia content with text. Without annotations, 
the file-name of a particular photo or video content is used as a default title. 
Unlike Twitter, Flicker imposes no size limit on the annotation text. Using 
Flickr to post (or view restricted) content, generally requires having an account.
However, public content can be viewed by anyone. Flickr has a notion of 
a {\em contact}, akin to a friend or a connection on other OSNs. 

As follows from the above description, each of these three OSNs 
is quite distinct in its primary mission. This makes the problem of 
linking accounts across them particularly challenging. 

\section{Problem Setting}
\label{sec:problem-setting}
The author attribution problem can be informally defined as:
\begin{quote}
{\em Given a set of known authors $A_{known} = \{a_1, a_2, ..., a_n\}$, 
and an anonymous contribution $C$ (textual, non-textual or a mix of both),
find the most likely candidate author of $C$ among those in $A_{known}$.}
\end{quote}
In the OSN context, author attribution problem translates into finding the most 
likely candidate author of anonymous posts, i.e., the user who most likely 
generated these posts given his or her OSN profile. We refer to 
attribution of anonymous posts to a user account as {\em linking}.

As mentioned earlier, our goal is to study the author 
attribution problem (based on stylometry) across multiple OSNs. Basically, we assume that
some people have accounts in two OSNs and we want to link these accounts. 
We have $OSN_1$ and $OSN_2$  each with its own
set of accounts. We first remove from each $OSN$ accounts that do not have a 
match (authored by the same user) in the other $OSN$. This results in $R\_OSN_1$ and $R\_OSN_2$ that are reduced versions of $OSN_1$ and $OSN_2$, respectively. To make the problem more 
challenging and also more realistic, we pollute $R\_OSN_2$ by introducing 
additional $X$ randomly chosen  accounts that were originally in $OSN_2$. 
As a result, for each account in $R\_OSN_1$, there is a matching account in 
$R\_OSN_2$. We refer to the accounts in $R\_OSN_1$ as {\em unknown},
and those in $R\_OSN_2$ -- as {\em known}, accounts. 

Now, the problem is reduced to finding a matching model $M$, i.e., an 
author attribution technique, that links unknown accounts in $R\_OSN_1$ 
to known accounts in $R\_OSN_2$. Specifically, for each unknown account in 
$R\_OSN_1$, $M$ returns a list of all accounts in $R\_OSN_2$ sorted in 
decreasing order of probability of the correct match. Similar to prior work in 
\cite{mish-linkability}, we define Top-$K$ linkability ratio $LR$ of $M$ as 
the ratio of unknown accounts (accounts in $R\_OSN_1$) that have their 
correct matching account -- in $R\_OSN_2$ -- among the Top-$K$ accounts 
of their returned lists from $M$. Our goal boils down to finding a matching model 
that maximizes $LR$ with respect to $X$ and $K$.
We vary $X$ so that the total number of known accounts ranges from $100$ to $100,000$.
Furthermore, we vary $K$ among $1$, $10$ and $100$.  

\section{Dataset}
\label{sec:dataset}
We use the base dataset obtained (crawled) and used by Goga et al. 
\cite{goga2013exploiting}.
Encompassing users from Yelp, Twitter and Flickr, this dataset is gigantic, 
containing over $350,000,000$ tweets, $29,000,000$ Flickr posts and 
$1,000,000$ Yelp reviews. Its most important property is the ground truth of 
matching accounts: it provides a set of users who operate accounts in multiple OSNs.
In the rest of this section, we describe the data cleaning process and then 
provide more details regarding matching accounts.

\subsection{Data Cleaning}
\label{subsec:data-cleaning}
Our initial analysis of the base dataset revealed the existence of numerous users with 
very limited overall contributions. However, stylometric analysis is known to 
perform accurately in the context of highly prolific users. Some recent studies 
\cite{afroz2014doppelganger, mcdonald2012use, rao2000can} report achieving 
good linkability performance with at least $4,500$ words per author. Thus, we 
first need to cull users with lower overall contributed text.   
We also need to filter out contributions that did not originate with the target user,
since some OSNs (e.g., Twitter) allow users to repost (re-tweet) what other people 
have posted. This filtering helps us better capture users' own stylometric properties. 
Consequently, we filter out the following:
\begin{compactitem}
  \item Twitter re-tweets  -- including tweets preceded with ``rt''. (Some Twitter users 
  copy \& paste tweets and add ``rt'' to start of a new tweet, instead of using 
  Twitter's official re-tweet functionality. This often occurs because Twitter does not 
  allow re-tweeting if the author of the original tweet has a private profile.)
  \item URLs, since they typically have no relevance to a user's stylometric profile.
  \item User mentions, identified via ``@'' character followed by a user-name.
  \item Posts in languages other than English. 
  
  \noindent{\bf NOTE:} we refer to a single piece of content generated by a user 
  as a {\em post}. 
  It denotes: a tweet in Twitter, a review in Yelp, and a photo annotation text in Flickr.
\end{compactitem}
After filtering, we combine all remaining posts of users into a single body of text.
This corresponds to the union of Yelp reviews, Twitter tweets and Flicker photo annotations.
As the last step, we remove all users who have a cumulative word count of less than
$1,000$. We stress that this threshold of only $1,000$ words per author is significantly 
lower than that in previous studies, e.g., $4,500$ words in \cite{afroz2014doppelganger, 
mcdonald2012use, rao2000can}. Table \ref{table:dataset} presents some dataset statistics
before and after data cleaning. The most important difference is the evident 
increase in average number of posts per user after cleaning.

\begin{table*}[t]
\centering
\caption{Dataset statistics before ($Yelp$, $Twitter$, $Flickr$) and after ($Yelp'$, $Twitter'$, $Flickr'$) cleaning.}
\begin{tabular}{l|c|c|c|c|c|c|}
\cline{2-7}
\textbf{} & \textit{$Yelp$} & \textit{$Yelp'$} & \textit{$Twitter$} & \textit{$Twitter'$} & \textit{$Flickr$} & \textit{$Flickr'$} \\ \hline
\multicolumn{1}{|l|}{\textbf{Number of users}} & 62,788 & 9,348 & 693,866 & 263,680 & 228,735 & 10,800 \\ \hline
\multicolumn{1}{|l|}{\textbf{Number of posts}} & \multicolumn{1}{l|}{1,260,927} & \multicolumn{1}{l|}{1,135,912} & \multicolumn{1}{l|}{359,015,338} & \multicolumn{1}{l|}{320,071,427} & \multicolumn{1}{l|}{29,521,599} & \multicolumn{1}{l|}{9,497,133} \\ \hline
\multicolumn{1}{|l|}{\textbf{Average number of posts per user}} & 20 & 122 & 517 & 1,214 & 129 & 879 \\ \hline
\multicolumn{1}{|l|}{\textbf{Average number of words in a post}} & 136 & 139 & 12 & 9 & 6 & 11 \\ \hline
\end{tabular}
\label{table:dataset}
\end{table*}

\subsection{Matching Accounts}
Dataset includes a set of matching accounts that correspond to what we refer to as: {\em ground truth}.
This set links user-names from different OSNs. This information was collected in 
\cite{goga2013exploiting} using the ``Friend Finder'' functionality provided in OSNs. 
Friend Finder was run on input of a list of $10,000,000$ e-mail addresses 
using browser automation tools: Watir and Selenuim\footnote{Watir: {\url{https://github.com/watir/watir}}
and Selenium: {\url{http://www.seleniumhq.org/}}}. Then, the list of users registered 
with the given e-mail addresses was checked, in order to identify  
user-names registered to the same e-mail address, i.e. operated by the same person.
Table \ref{table:matching_accounts} shows the number of matching accounts in the 
original base dataset. We also present the number of matching accounts after data 
cleaning, as described above.

\begin{table}[h]
\caption{Number of matching accounts across OSNs. Since cleaning eliminates all
non-English and low-contribution accounts, the number of matching accounts decreases 
notably. However, a sufficient number of matching accounts remain for linkability 
experiments.}
\centering
\begin{tabular}{l|c|c|}
\cline{2-3}
 & \textit{\begin{tabular}[c]{@{}c@{}}Original Dataset\end{tabular}} & \textit{\begin{tabular}[c]{@{}c@{}}Cleaning w/ words $\geq$ 1000\end{tabular}} \\ \hline
\multicolumn{1}{|l|}{\textbf{Yelp-Twitter}} & 1,889 & 153 \\ \hline
\multicolumn{1}{|l|}{\textbf{Twitter-Flickr}} & 13,629 & 299 \\ \hline
\multicolumn{1}{|l|}{\textbf{Yelp-Flickr}} & 1,199 & 55 \\ \hline
\end{tabular}
\label{table:matching_accounts}
\end{table}

\section{Preliminaries}
\label{sec:preliminaries}
Before presenting experimental results, we provide some background
information about the feature set and the methodology.

\subsection{Feature Set}
\label{subsec:features}
We construct an unique set of features, using a subset of the popular 
Writeprints set \cite{abbasi2008writeprints} along with $3$ additional features.  
Writeprints contains $22$ distinct stylometric features. From these, we select 
$9$ lexical, syntactic and content features before adding $3$ more custom
features (not present in Writeprints). The resulting $12$ features are:
\begin{compactitem}
\item \textbf{Lexical} features include frequencies of alphabetical n-grams 
(n consecutive letters) and special characters, e.g. ``*", ``@'', ``\#'', ``\$'', 
and ``\%''.
\item \textbf{Syntactic} features consist of frequencies of function words, punctuation 
characters and Part-Of-Speech (POS) tags, where unigrams correspond to one tag,
and bigrams to two consecutive tags. Function words are 512 common 
function words used by Koppel et al. \cite{koppel2005automatically}.

POS tags are grammatical descriptions of words in sentences, e.g, adjective,
noun, verb and adverb. We use two popular POS taggers:
\begin{enumerate}
  \item Stanford Log-linear \cite{toutanova2003feature}, which was the booster of
  account linkability in recent studies \cite{afroz2014doppelganger,almishari2014fighting}.
  \item GATE Twitter \cite{derczynski2013twitter}, which has never been used in
  account linkability before.
\end{enumerate}
POS tagging of tweets is hard due to the short message style in Twitter.
Therefore, we integrate GATE -- a state-of-art accurate POS tagger specially designed for Twitter -- to our feature set.
Our experimental results demonstrate that GATE Twitter tagger improves the account
linkability significantly.
\item \textbf{Content} features include frequency of words. This is the only stylometric 
feature used in \cite{goga2013exploiting} for linking accounts.
\end{compactitem}
Table \ref{table:features} lists our feature categories and the number of 
features within each. Features are computed for each user profile. 
Each feature is normalized by the total count of features within the same category.

Similar subsets of Writeprints were used in several prior linkability studies, e.g., 
Afroz et al. \cite{afroz2014doppelganger,almishari2014fighting}, to yield high 
linkability accuracy. Encouraging results using Letter Quads (4-grams) are achieved in 
Kevselj et al. \cite{kevselj2003n}.  To the best of our knowledge,
GATE Twitter POS features have never been used in linkability studies before.

\begin{table}[h]
\begin{center}
\caption{Lexical, syntactic and content features in our feature set. Boldfaced features are not
in Writeprints.}
\begin{tabular}{| l | ll |}
\hline
\textbf{Features}             & \textbf{Count}       &  \\ \hline\cline{1-3}
Letter n-grams, $n=1,2,3$         & $26^n$ &  \\ \hline
{\bf Letter 4-grams} & $26^4$ &  \\ \hline
Special characters           & $20$                 &  \\ \hline
Function words               & Dynamic              &  \\ \hline
Punctuation marks            & $8$                  &  \\ \hline
Stanford POS n-grams, n=1,2 & Dynamic              &  \\ \hline
{\bf Gate POS n-grams, n=1,2} & Dynamic              &  \\ \hline
Words                        & Dynamic              &  \\ \hline
\end{tabular}
\label{table:features}
\end{center}
\end{table}

\subsection{Methodology}
\label{subsec:methodology}
Based on the setting described in Section \ref{sec:problem-setting}, we have two 
sets of accounts: known and unknown. We want to accurately match unknown 
accounts to their known counterparts, while maintaining the highest possible Top-$K$  
Linkability Ratio (LR). For that, we first convert each user profile into a 
feature vector: $F_T=\{F_{T_1}, F_{T_2}, ..., F_{T_n}\}$ where $F_{T_i}$ 
denotes the $i$-th token for feature $F_T$.

Next, we initiate a distance learning model using Chi-Square Distance ($CS_d$)  
to link an unknown account to a known one. Specifically, for each unknown 
account $a_u$, we calculate the $CS_d(a_u,a_{k_j})$ where $j$ varies over all 
possible known accounts. Finally, we rank the distances in ascending order 
and output the resulting ordered list, where the first entry represents the most 
likely match of the known account $a_k$ to the unknown account $a_u$.

\section{Experimental Results}
\label{sec:results}
This section presents the results of the large-scale trilateral OSN account
linkability study. We  begin with the baseline result. Next, we outline the new
Multi-Level Linker Framework which significantly  improves on the baseline.
Then, we show how this framework yields scalable linkability ratios (LRs) 
for up to $100,000$ authors.
Finally, we present and discuss experiment execution times \& memory footprint.

\subsection{Baseline}
\label{subsec:baseline}
Using the methodology from the previous section, we experiment with various
features. Similar to prior work in \cite{almishari2014fighting},
we apply a greedy hill-climbing algorithm to assess the effects of every feature.
We start with all features individually. Then, we combine the best-performing 
features and assess the amount of improvement. We present the baseline assessment 
only for Yelp$\leftrightarrow$Twitter linkability, since other sets perform similarly.
Following Section \ref{sec:problem-setting}, we set the list of unknown accounts $A_{unknown}$ to  the full-set of matching accounts as (153 accounts) while we set the size of the known accounts $A_{known}$ to $1000$ accounts.

Table \ref{table:baseline-single} shows Top-1 LRs of individual features. At best,
Yelp$\rightarrow$Twitter already shows a relatively high 
55\% Top-1 LR, while Twitter$\rightarrow$Yelp performs quite poorly, at 10\%.

\begin{table}[h]
\caption{Top-1 LRs using the baseline Chi-Square methodology.
Boldfaced cells represent the highest LRs.}
\centering
\begin{tabular}{|l|c|c|}
\hline
\textbf{Feature Index} & \textbf{Twitter$\rightarrow$Yelp} & \textbf{Yelp$\rightarrow$Twitter} 
\\ \hline \cline{1-3}
\textsf{1: Letter Uni} & 1\% & 1\% \\ \hline
\textsf{2: Letter Bi} & 1\% & 43\% \\ \hline
\textsf{\bf{3: Letter Tri}} & {\bf 7\%} & {\bf 55\%} \\ \hline
\textsf{\bf{4: Letter Quad}} & {\bf 10\%} & {\bf 53\%} \\ \hline
\textsf{5: Special Chars} & 1\% & 0\% \\ \hline
\textsf{\bf{6: Func. Words}} & 3\% & {\bf 50\%} \\ \hline
\textsf{7: Punc. Marks} & 0\% & 1\% \\ \hline
\textsf{8: Stanford POS Tags Uni} & 1\% & 8\% \\ \hline
\textsf{9: Stanford POS Tags Bi} & 3\% & 27\% \\ \hline
\textsf{\bf{10: Words}} & {\bf 9\%} & 39\% \\ \hline
\textsf{11: GATE POS Tags Uni} & 2\% & 7\% \\ \hline
\textsf{12: GATE POS Tags Bi} & 3\% & 18\% \\ \hline
\end{tabular}
\label{table:baseline-single}
\end{table}

Next, we combine the best features (highlighted in boldface) from 
Table \ref{table:baseline-single} and show improved results in 
Table \ref{table:baseline-combined}.

\begin{table}[h]
\caption{Top-1 LRs, with combined best features from Table \ref{table:baseline-single}.}
\centering
\begin{tabular}{|l|c|c|c|c|}
\hline
Features & 4\&10 & 3\&10 & 3\&4 & 3\&4\&10 \\ \hline\cline{1-5}
Twitter$\rightarrow$Yelp & \textbf{11\%} & \textbf{8\%} & \textbf{9\%} & \textbf{9\%} \\ \hline 
\multicolumn{5}{c}{}\\ \hline
Features & 3\&4 & 3\&6 & 4\&6 & 3\&4\&6 \\ \hline \cline{1-5}
Yelp$\rightarrow$Twitter & \textbf{54\%} & \textbf{59\%} & \textbf{57\%} & \textbf{56\%} \\ \hline
\end{tabular}
\label{table:baseline-combined}
\end{table}

%
%
\begin{figure*}[b]
  \centering
  \subfigure[OSN$\rightarrow$Twitter]{
    \label{fig:easy-to-twitter}
    \includegraphics[width=0.47\textwidth,height=6.5cm]{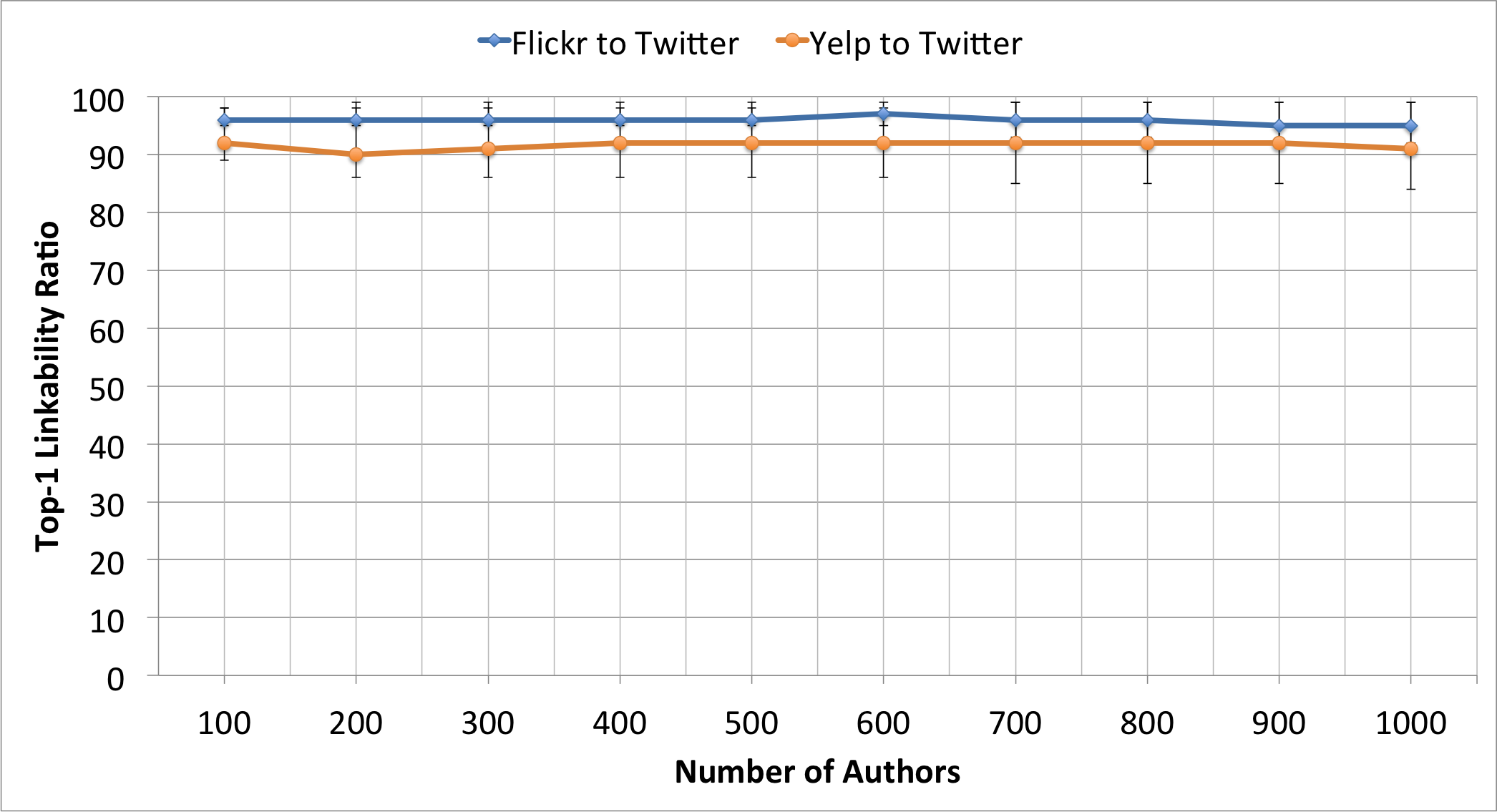}}
  \subfigure[OSN$\rightarrow$Yelp]{
    \label{fig:easy-to-yelp}
    \includegraphics[width=0.47\textwidth,height=6.5cm]{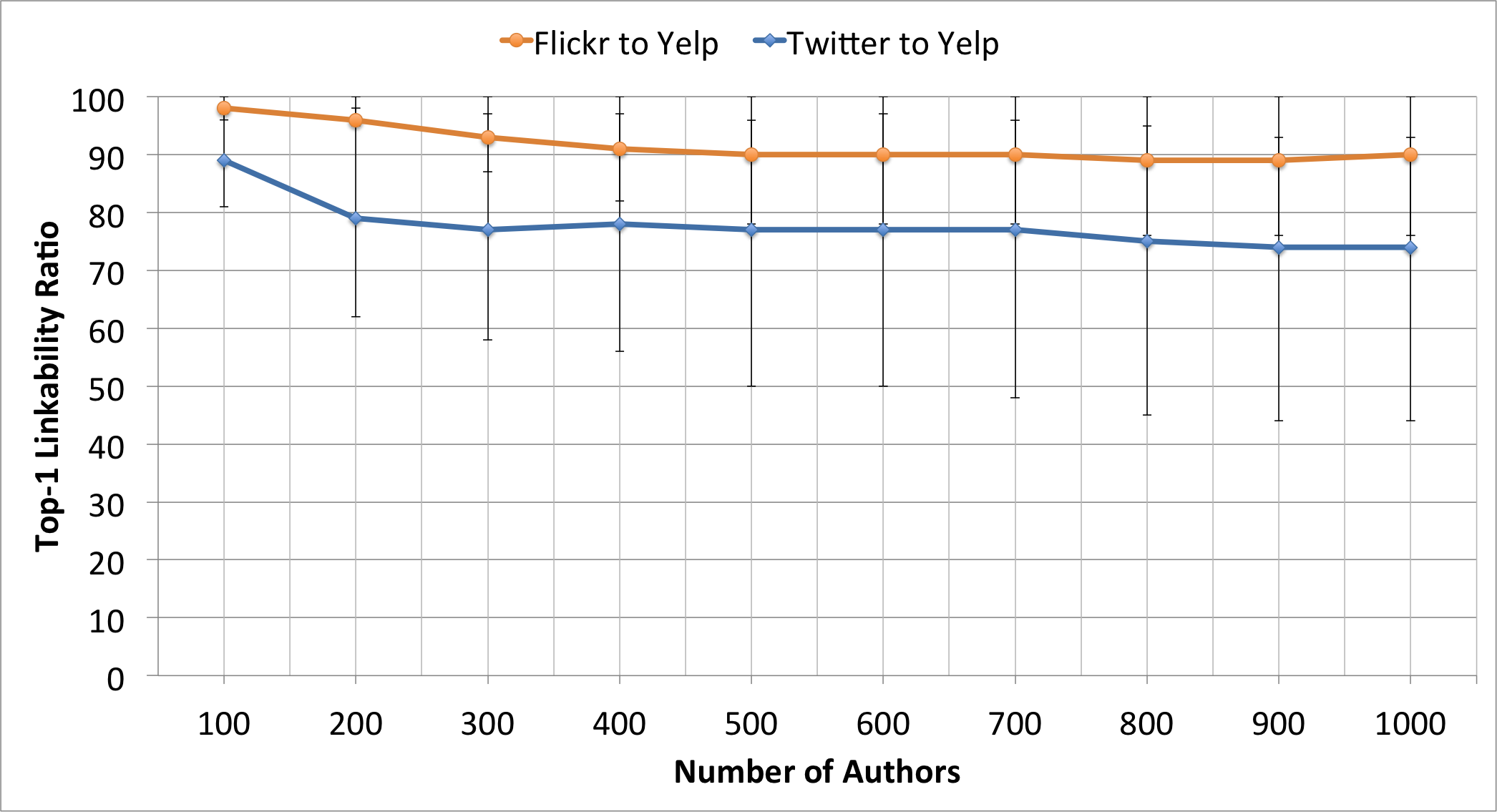}}
  \caption{Top-1 LRs when number of authors is increased from 100 to 1,000}
  \label{fig:easy}
\end{figure*}

For the Twitter$\rightarrow$Yelp case, when Letter Quadgrams and
Words features are combined, results are slightly better than the baseline. 
However, after combining more than two features, we observe a decrease in LR.
As for Yelp$\rightarrow$Twitter, LR increases slightly when best 
features are combined (3\&6). Similar to Twitter$\rightarrow$Yelp, 
combining more than two features decreases LR.

These results are comparable to those obtained 
in language-style correlation investigated in \cite{goga2013exploiting}. 
Likewise, we achieve modest LRs, even with more complex language-based features.
To summarize, recent techniques that work reasonably well within the same OSNs, 
do not appear to be as effective across OSNs.
To this end, in the next section, we construct the Multi-Level Linker Framework,
which, according to our experiments, significantly boosts linkability.

\subsection{Multi-Level Linker Framework (MLLF)}
\label{subsec:framework}
While experimenting with various combination of features, we notice that combining
many of them increases noise and prolongs run-times. Moreover, dimensionality 
reduction techniques like SVD, do not help increase linkability. This motivates us to 
explore how to make better use of all textual features.

We now present the Multi-Level Linker Framework (MLLF). The intuition behind it
is the use of features in a more hierarchical manner. The basic idea is to run 
linkability experiments at multiple levels, with each level using a different feature category.
After each level, we halve the number of known authors, for every unknown author.
This is done by filtering out the most distant (least likely) known authors.
Then, at the next level, we use a different feature category with the most likely known authors.
We apply this technique for every feature category, and eventually output the final 
linkability -- the final position of the matching account.
In every experiment, we randomly permute the order of feature categories.
We run experiments in $10$-fold and report the averages of final linkability results.
In plots, we provide positive and negative error bars to average linkability
results in order to better understand the effects of feature ordering.
High level pseudo-code of MLLF can be found in the Appendix. 

Applying MLLF yields significantly higher LRs with respect to the baseline.
Improvements -- between [27\%, 73\%]  -- in Top-$1$ LR, when the number 
of known authors is $1,000$, are:

\begin{table}[h]
\centering
\begin{tabular}{| c | c |} 
\hline
  Twitter$\rightarrow$Yelp & 11\% $\rightarrow$ 63\% \\ \hline
  Yelp$\rightarrow$Twitter & 59\% $\rightarrow$ 88\% 
  \\ \hline
  Twitter$\rightarrow$Flickr & 11\% $\rightarrow$ 54\%
   \\ \hline
  Flickr$\rightarrow$Twitter & 67\% $\rightarrow$ 94\%
   \\ \hline
  Flickr$\rightarrow$Yelp & 13\% $\rightarrow$ 86\%
   \\ \hline
  Yelp$\rightarrow$Flickr & 5\% $\rightarrow$ 66\%
   \\ \hline
\end{tabular}
\end{table}

\subsection{Scalability: Number of Authors}
\label{subsec:scalability}
Having obtained an improvement over baseline results, we now  
consider MLLF's scalability. To this end, we vary the number 
of known authors from $100$ to $100,000$ and examine how LRs 
are affected.

\subsubsection{From $100$ to $1,000$}
\label{subsec:easy}
In the first batch, we experiment with $|A_{known}|$ from $100$ to $1,000$.
OSN pairs with the highest Top-$1$ LRs are shown in Figure \ref{fig:easy}.
OSN$\rightarrow$Twitter LRs gets as high as 95\% (Figure \ref{fig:easy-to-twitter})
while OSN$\rightarrow$Yelp LRs gets 90\% (Figure \ref{fig:easy-to-yelp}) in a set of $1,000$ authors.
We notice linkability to Twitter is higher than linkability to Yelp in all cases.
Also, when number of author increases, OSN$\rightarrow$Yelp LRs decreases
more than OSN$\rightarrow$Twitter.
Lastly, OSN$\rightarrow$Yelp linkability results shows higher variance, that is
affected more by the order features.

%
%
\begin{figure*}[t]
  \centering
  \includegraphics[width=0.47\textwidth,height=6.5cm]{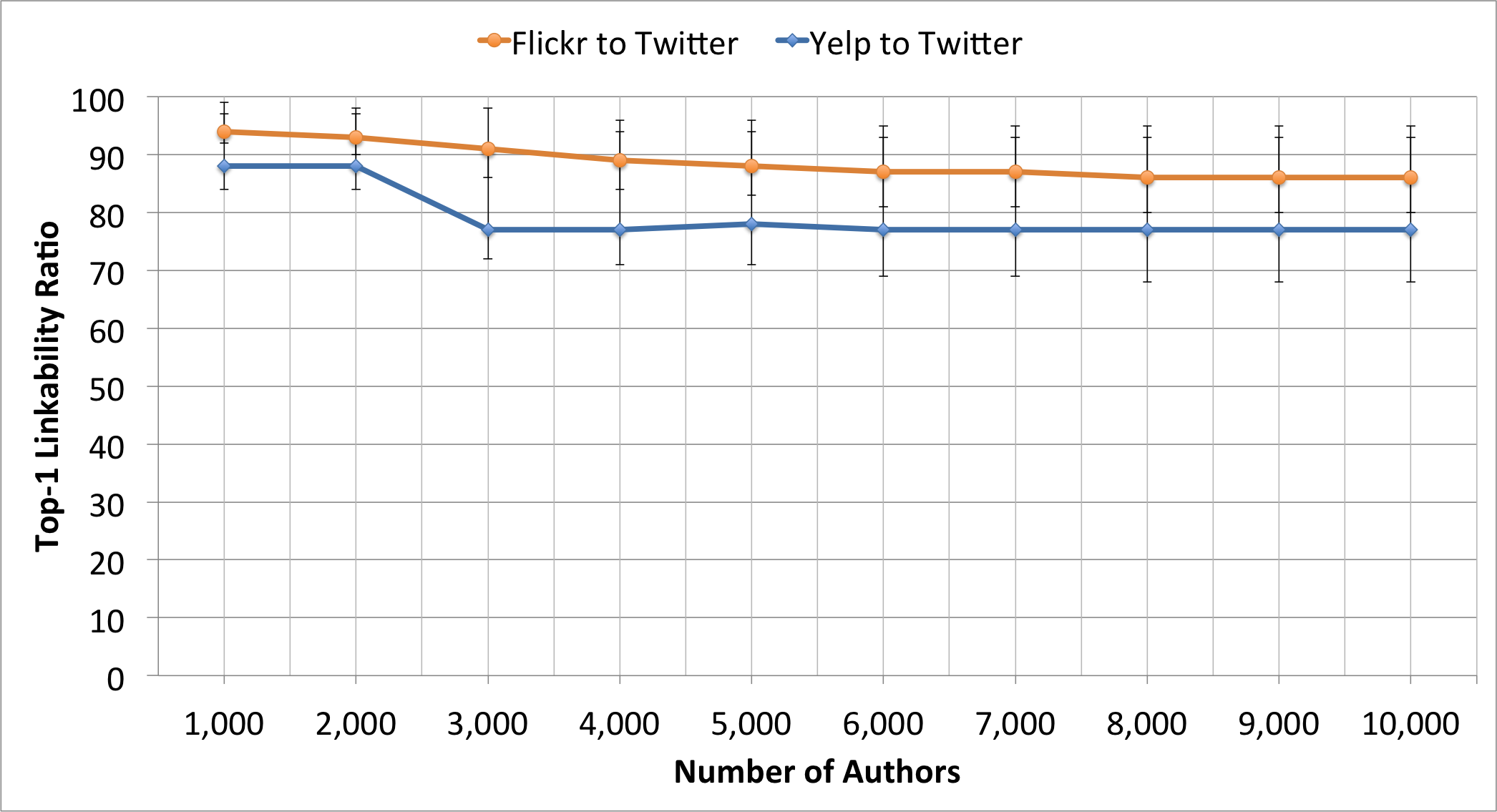}
  \caption{Top-$1$ LRs of OSN$\rightarrow$Twitter when number of authors grows from $1,000$ to $10,000$}
  \label{fig:medium-to-twitter}
\end{figure*}

%
%
\begin{figure*}[t]
  \centering
  \subfigure[OSN$\rightarrow$Yelp]{
    \label{fig:medium-to-yelp}
    \includegraphics[width=0.47\textwidth,height=6.5cm]{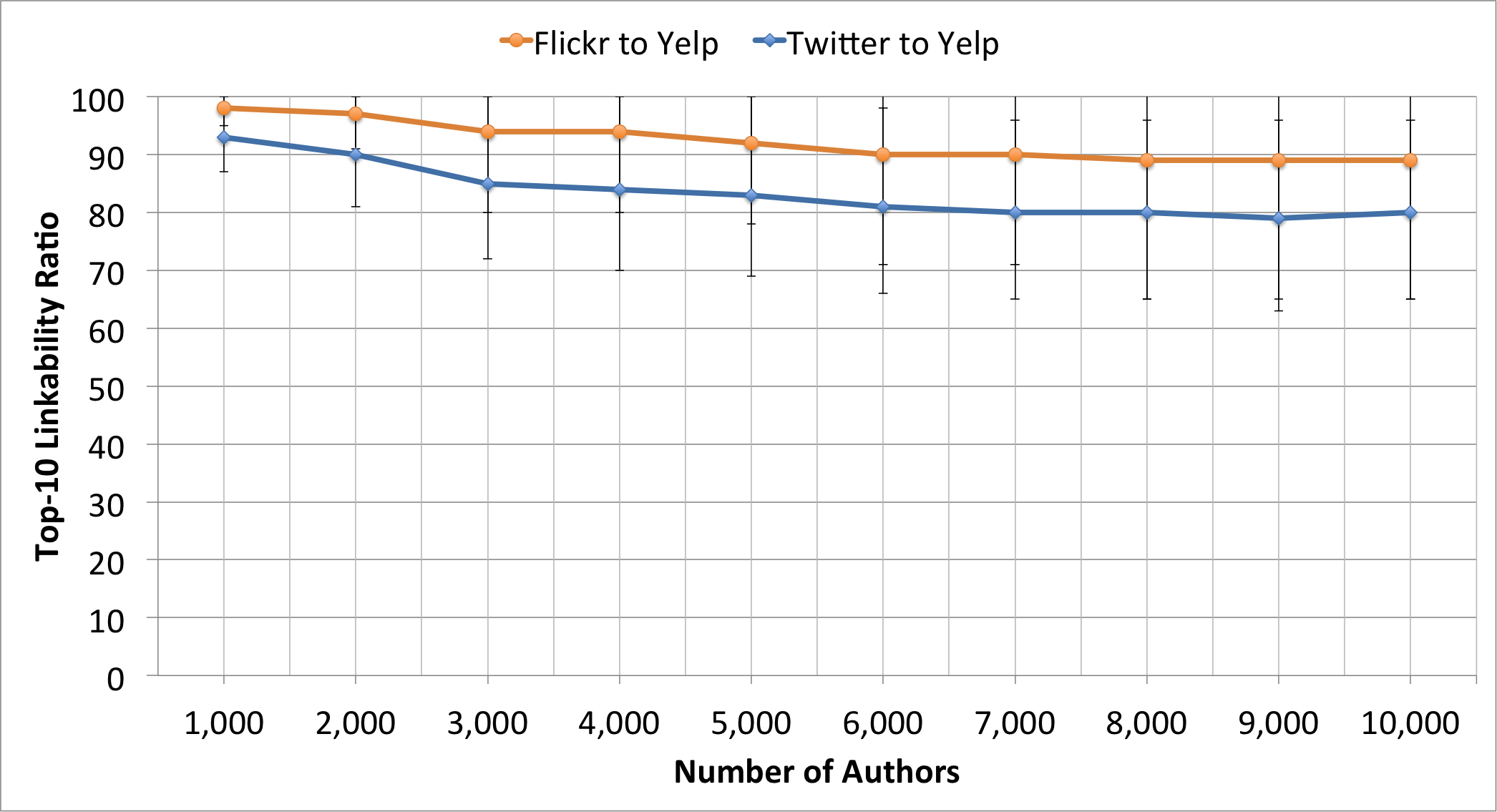}}
  \subfigure[OSN$\rightarrow$Flickr]{
    \label{fig:medium-to-flickr}
    \includegraphics[width=0.47\textwidth,height=6.5cm]{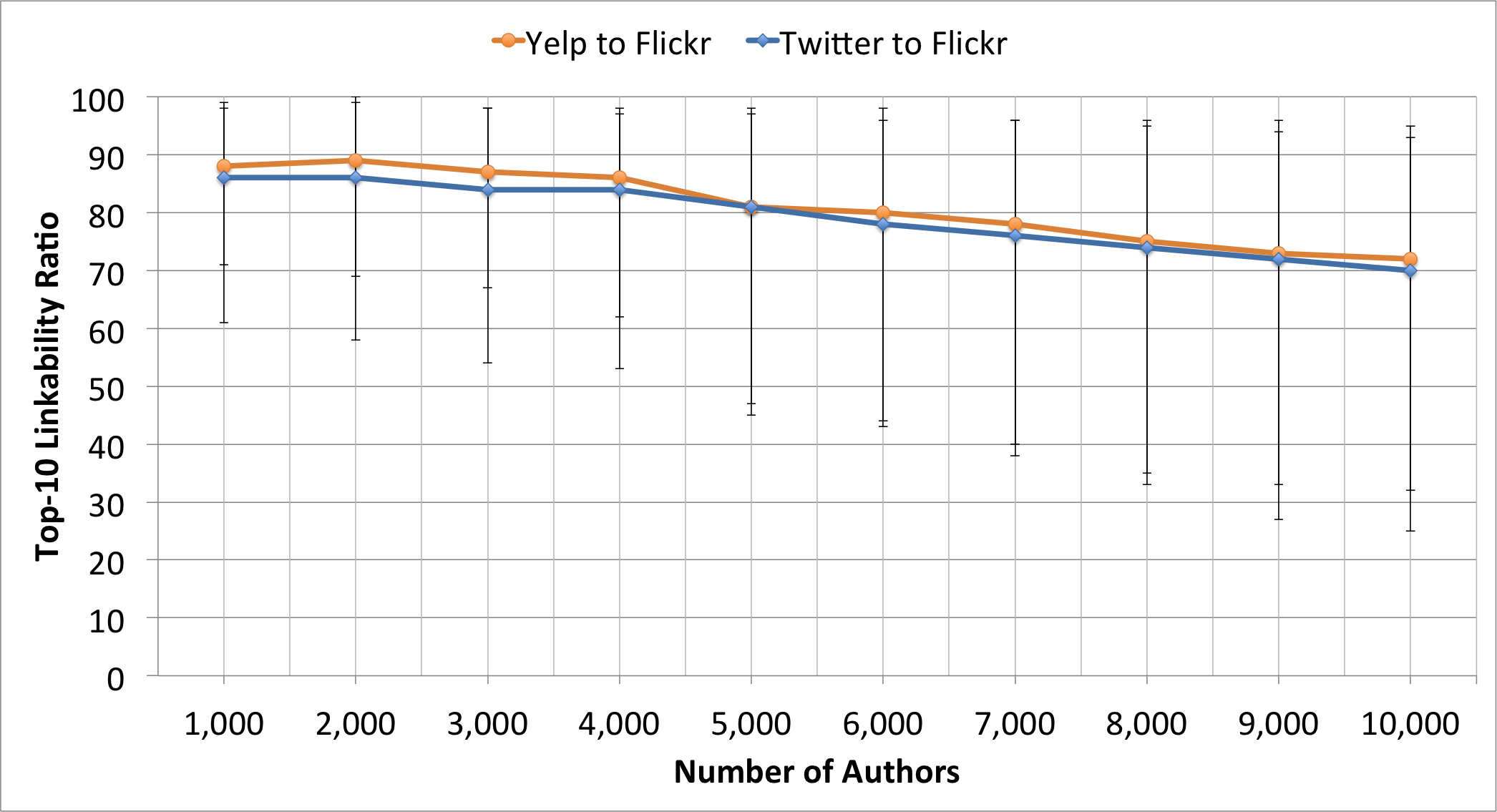}}
  \caption{Top-$10$ LRs when number of authors grows from $1,000$ to $10,000$}
  \label{fig:medium-top10}
\end{figure*}

OSN$\rightarrow$Flickr exhibits the worst results; LRs are shown in Table \ref{table:easy}.
Top-$1$ LR of Twitter$\rightarrow$Flickr drops to 63\% in a set of $1,000$ authors.
Interestingly, LRs of OSN$\rightarrow$Flickr does not decrease as much as OSN$\rightarrow$Yelp.
While Top-$1$ LRs of OSN$\rightarrow$Yelp decreases as much as 15\%, the
biggest decrease is only 4\% for OSN$\rightarrow$Flickr when number of authors grows
from $100$ to $1,000$.

\begin{table}[h]
\caption{Top-1 and Top-10 LRs of OSN$\rightarrow$Flickr as the number of authors grows from $100$ to $1,000$}
\centering
\begin{tabular}{l|c|c|c|c|}
\cline{2-5}
& \multicolumn{2}{c|}{\textit{Top-1}} & \multicolumn{2}{c|}{\textit{Top-10}} \\ \hline
\multicolumn{1}{|l|}{\textbf{Number of Authors}} & \textit{100} & \textit{1,000} & 
\textit{100} & \textit{1,000} \\ \hline
\multicolumn{1}{|l|}{\textbf{Yelp$\rightarrow$Flickr}} & 77\% & 73\% & 93\% & 92\% \\ \hline
\multicolumn{1}{|l|}{\textbf{Twitter$\rightarrow$Flickr}} & 65\% & 63\% & 88\% & 89\% \\ \hline
\end{tabular}
\label{table:easy}
\end{table}

\subsubsection{From $1,000$ to $10,000$}
Next, we vary the number of authors from $1,000$ to $10,000$.
(The actual number of accounts in $Yelp'$ is  $9,348$, which we round toа
$10,000$ to simplify the graphs.)

Firstly, we show Top-$1$ LRs of OSN$\rightarrow$Twitter in Figure \ref{fig:medium-to-twitter}.
Similar to trends in Section \ref{subsec:easy}, the highest Top-$1$ LRs among all OSN
combinations is 86\% for Flickr$\rightarrow$Twitter, followed by 77\% for Yelp$\rightarrow$Twitter
when the number of authors is $10,000$.
Moreover, OSN$\rightarrow$Twitter model continues to show low linkability variance
-- 6\% in Flickr$\rightarrow$Twitter and 9\% in Yelp$\rightarrow$Twitter --
according to the order of features.

Secondly, Top-$10$ LRs of OSN$\rightarrow$Yelp are shown Figure \ref{fig:medium-to-yelp}
and OSN$\rightarrow$Flickr in Figure \ref{fig:medium-to-flickr}.
OSN$\rightarrow$Yelp and OSN$\rightarrow$Flickr perform worse than OSN$\rightarrow$Twitter.
Therefore, we present the graphs in Top-$10$ and show Top-$1$ values in Table \ref{table:medium}.
We observe that Flickr$\rightarrow$Yelp LR achieves 89\% and Twitter$\rightarrow$Yelp achieves 
80\% in Top-$10$. In contrast, Yelp$\rightarrow$Flickr is 72\% and Twitter$\rightarrow$Flickr is 70\%.
Also, OSN$\rightarrow$Yelp is more resilient to random feature ordering than OSN$\rightarrow$Flickr.
Furthermore, both Yelp and Twitter perform very similarly when linking to a Flickr account.

Finally, Table \ref{table:medium} summarizes linkability results for all OSN 
combinations. Top-$1$ LR for $10,000$ authors drops to as low as 
29\% in Yelp$\rightarrow$Flickr, and grows as high as 86\% in Flickr$\rightarrow$Twitter.
For Top-$1$, linkability to Twitter is best, while linkability to Flickr is worst.
For Top-$10$, the results are really encouraging with 70\% as the lowest LR for 
a set of $10,000$ authors. Lastly, linkability to Twitter decreases by only 
2\% when number of authors changes from $1,000$ to $10,000$.

%
%
\begin{figure*}[t]
  \centering
  \subfigure[From 10,000 to 100,000; Top-$100$]{
    \label{fig:hard}
    \includegraphics[width=0.47\textwidth,height=6.5cm]{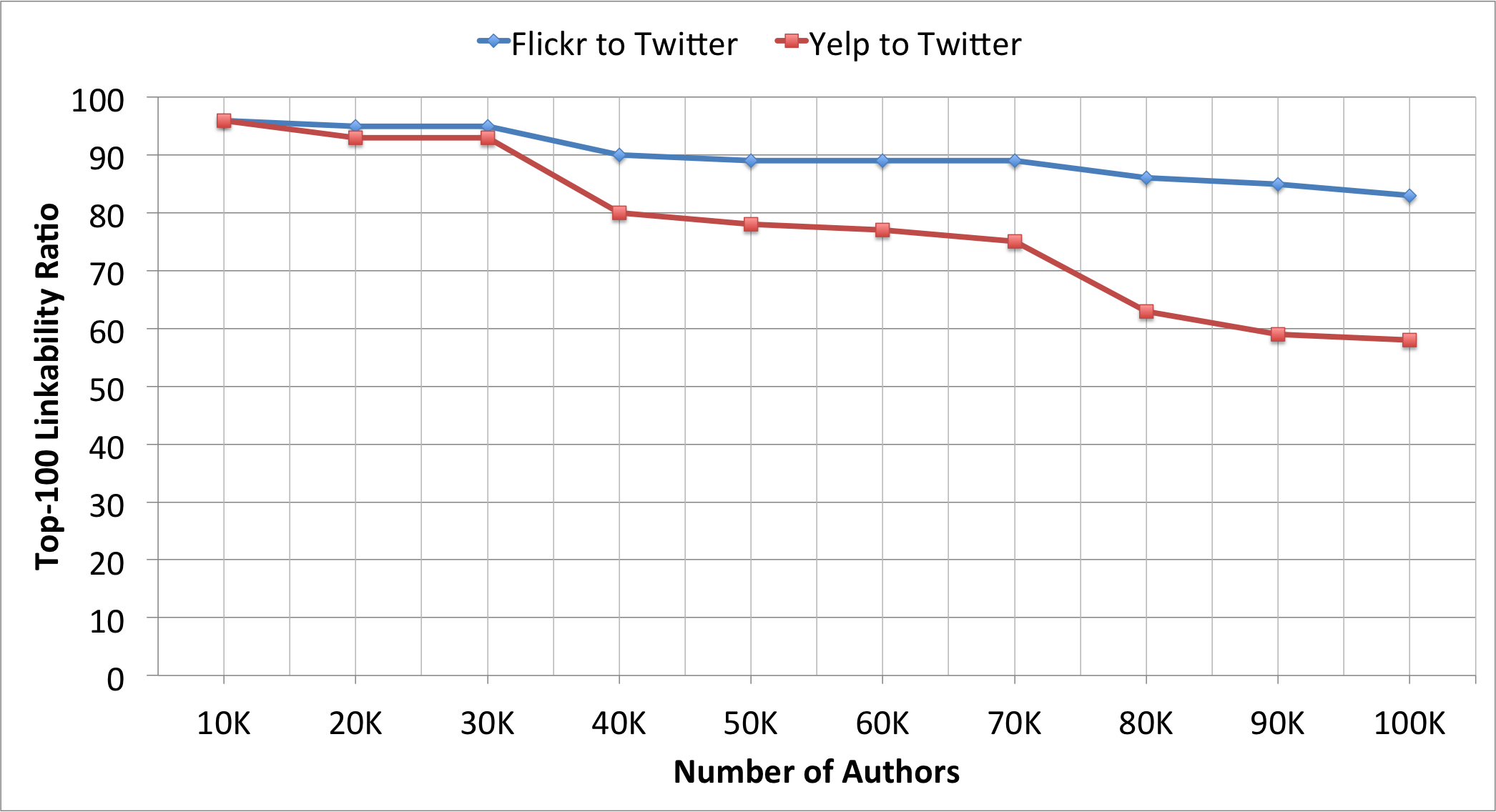}}
  \subfigure[From 100 to 100,000; Top-$10$]{
    \label{fig:combined_top10}
    \includegraphics[width=0.47\textwidth,height=6.5cm]{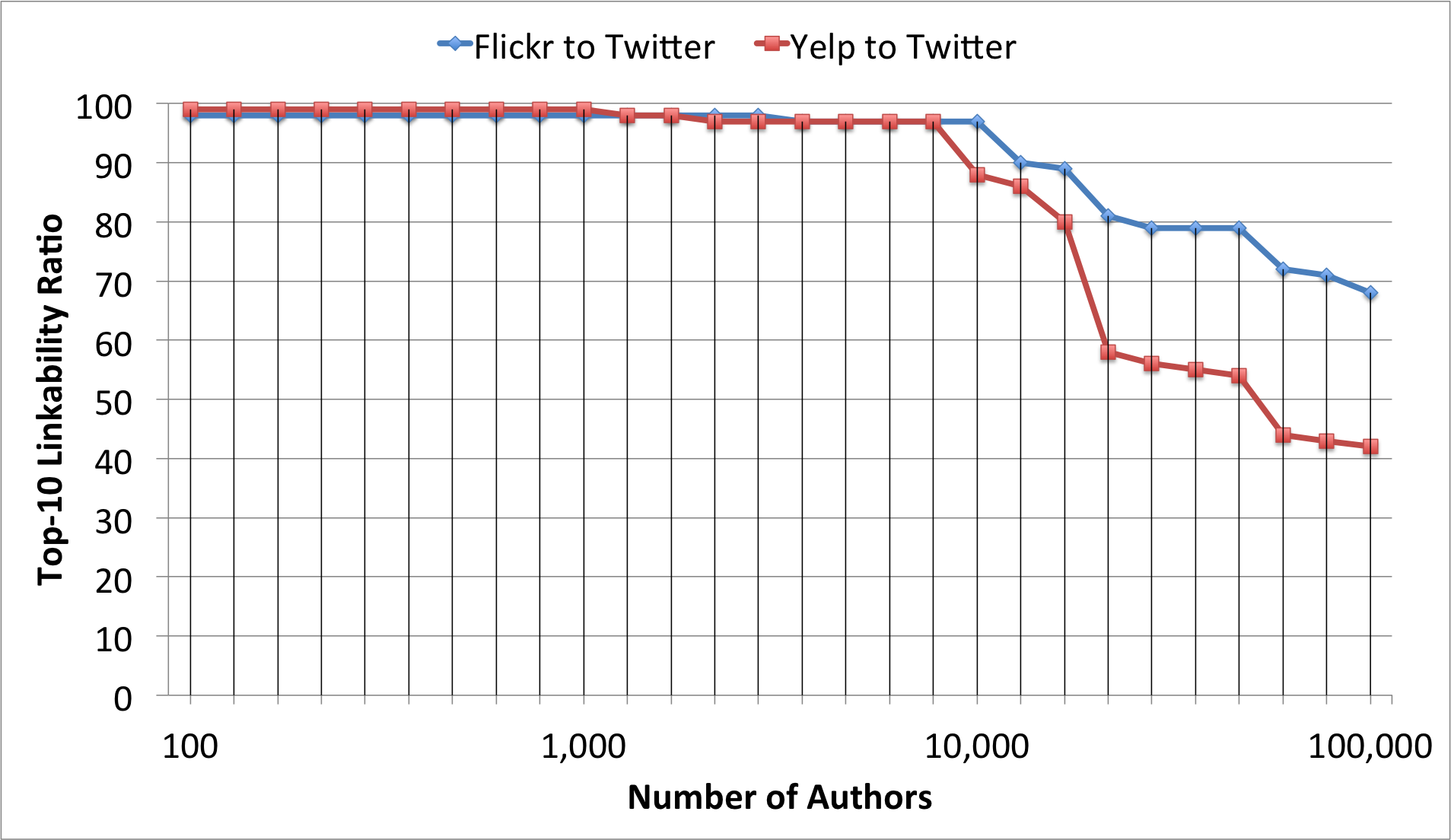}}
  \caption{LRs when \# of authors grows from $100$ to $100,000$}
  \label{fig:hard-main}
\end{figure*}

\begin{table}[h]
\caption{Top-1 and Top-10 LRs when \# of authors grows from $1,000$ to $10,000$}
\centering
\begin{tabular}{l|c|c|c|c|}
\cline{2-5}
 & \multicolumn{2}{c|}{\textit{Top-1}} & \multicolumn{2}{c|}{\textit{Top-10}} \\ \hline
\multicolumn{1}{|l|}{\textbf{Number of Authors}} & \textit{1,000} & \textit{10,000} & \textit{1,000} & \textit{10,000} \\ \hline
\multicolumn{1}{|l|}{\textbf{Flickr$\rightarrow$Twitter}} & 94\% & 86\% & 98\% & 97\% \\ \hline
\multicolumn{1}{|l|}{\textbf{Yelp$\rightarrow$Twitter}} & 88\% & 77\% & 99\% & 97\% \\ \hline
\multicolumn{1}{|l|}{\textbf{Flickr$\rightarrow$Yelp}} & 86\% & 63\% & 98\% & 89\% \\ \hline
\multicolumn{1}{|l|}{\textbf{Twitter$\rightarrow$Yelp}} & 63\% & 45\% & 93\% & 80\% \\ \hline
\multicolumn{1}{|l|}{\textbf{Yelp$\rightarrow$Flickr}} & 66\% & 29\% & 88\% & 72\% \\ \hline
\multicolumn{1}{|l|}{\textbf{Twitter$\rightarrow$Flickr}} & 54\% & 38\% & 86\% & 70\% \\ \hline
\end{tabular}
\label{table:medium}
\end{table}

\subsubsection{From $10,000$ to $100,000$}
\label{subsec:hard}
As the final step in the scalability exercise, we increase $|A_{known}|$ to $100,000$ authors.
As evident from Table \ref{table:dataset}, only $Twitter$ has up to $100,000$ authors after 
cleaning. Thus, we only experiment with Flickr$\rightarrow$Twitter and Yelp$\rightarrow$Twitter
combinations. Also, we remove Letter Quadgrams from the feature set and run this batch of 
experiments with the remaining $11$ features, due to memory problems experienced 
with over $90,000$ authors.

Figure \ref{fig:hard} shows Top-$100$ LRs and Table \ref{table:authors-10000-to-100000} shows 
Top-$1$ and Top-$10$ LRs. Notably, even in the extreme case of $100,000$ authors, 
we can still link to the known author with 54\% accuracy in Flickr$\rightarrow$Twitter, and 
18\% accuracy in Yelp$\rightarrow$Twitter. If we relax the linkability goal to Top-$100$,
Flickr$\rightarrow$Twitter grows to 83\% and Yelp$\rightarrow$Twitter to 58\%.
We notice that linkability from Flickr is higher than that from Yelp.
Moreover, the former is less affected by the increase in the number of authors: 
Flickr$\rightarrow$Twitter Top-$1$ LR decreases by 26\% while Yelp$\rightarrow$Twitter decreases by 50\%.

We also demonstrate Top-$10$ LRs from $100$ to $100,000$ authors in 
Figure \ref{fig:combined_top10}. We observe a decrease in LRs after $9,000$ authors,
and a sharp fall after $40,000$ authors. However, we still find our results
highly encouraging, and scary for authorship privacy, since we are only reporting Top-$10$
LRs, 0.01\% of all possible authors.

Our results significantly improve on the prior work of Goga, et al \cite{goga2013exploiting}.
Even though their setting is slightly different from ours,
we achieve True Positive Rate of 60\% in Flickr$\rightarrow$Twitter
and 36\% in Yelp$\rightarrow$Twitter in a set of 70,000 authors,
while \cite{goga2013exploiting} reaches 13\% for the former and 9\% for the latter
using language profile in a set of $75,747$ authors.\footnote{We set Top-$1$ LRs as True Positive Rate.}
As a similar result, both our and \cite{goga2013exploiting}'s experiments show
that linkability of Flickr$\rightarrow$Twitter is higher than Yelp$\rightarrow$Twitter.
We discuss some possible reasons in Section \ref{sec:discussion}.
Finally, since \cite{goga2013exploiting} did not experiment with any other OSN pairs,
we cannot compare our other linkability ratios.

%
%
\begin{figure*}[t]
  \centering
  \subfigure[From 100 to 10,000]{
    \label{fig:execution-easy-medium}
    \includegraphics[width=0.47\textwidth,height=6cm]{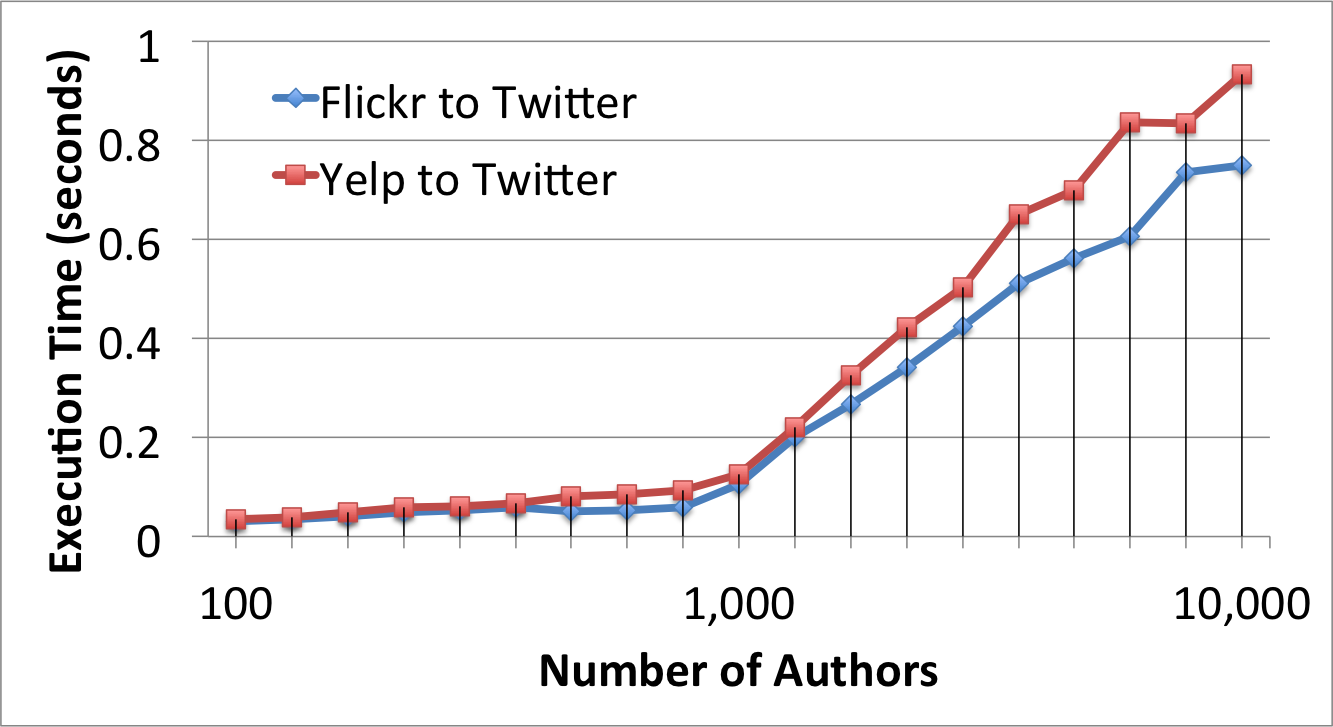}}
  \subfigure[From 10,000 to 100,000]{
    \label{fig:execution-hard}
    \includegraphics[width=0.47\textwidth,height=6cm]{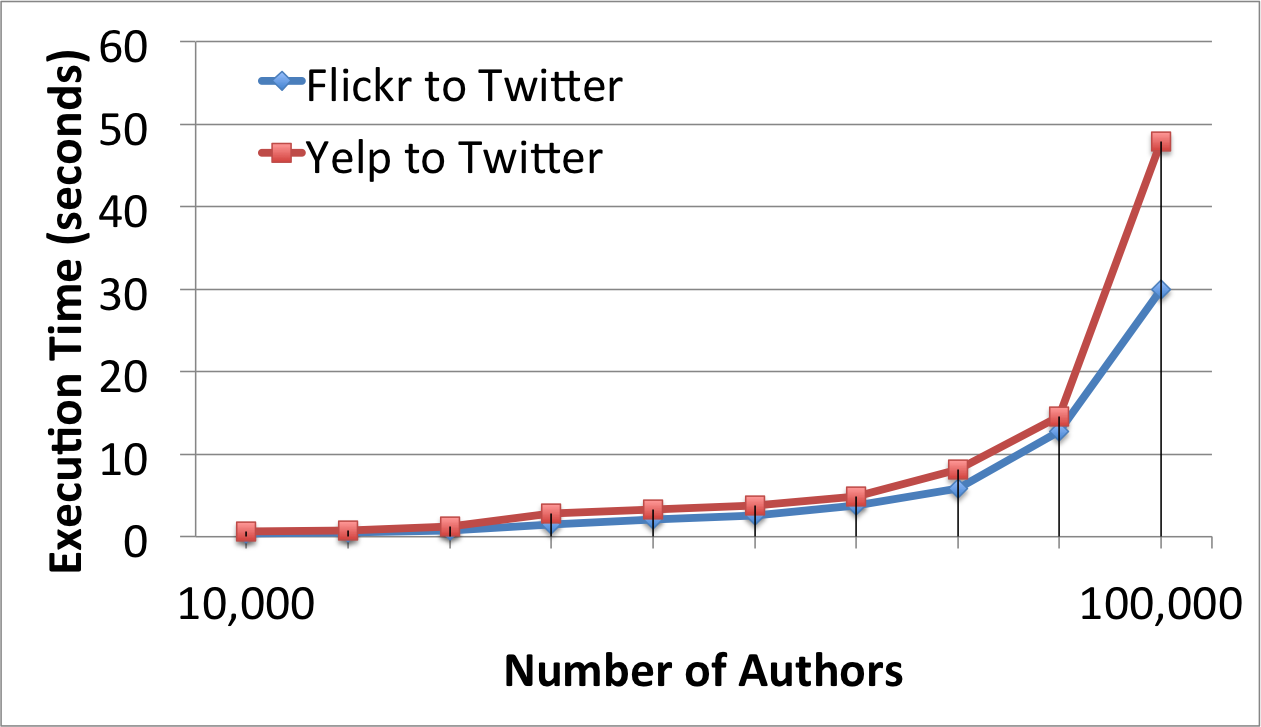}}
  \caption{Execution times of MLLF with variable \# of authors}
  \label{fig:execution}
\end{figure*}

\begin{table}[h]
\caption{Top-1 and Top-10 LRs as \# of authors grows from $10,000$ to $100,000$.}
\centering
\begin{tabular}{l|c|c|c|c|}
\cline{2-5}
 & \multicolumn{2}{c|}{\textit{Top-1}} & \multicolumn{2}{c|}{\textit{Top-10}} \\ \hline
\multicolumn{1}{|l|}{\textit{Number of Authors}} & \textit{10,000} & \textit{100,000} & \textit{10,000} & \textit{100,000} \\ \hline
\multicolumn{1}{|l|}{\textbf{Flickr$\rightarrow$Twitter}} & 80\% & 54\% & 91\% & 68\% \\ \hline
\multicolumn{1}{|l|}{\textbf{Yelp$\rightarrow$Twitter}} & 68\% & 18\% & 88\% & 42\% \\ \hline
\end{tabular}
\label{table:authors-10000-to-100000}
\end{table}

\subsection{Execution Time and Memory Footprint}
\label{subsec:execution-time}
Scalability in real-world OSNs begins with at least several millions of users.
Therefore, it is very important to assess performance of a linkability study
(such as ours) in order to test whether it is applicable in the real world.

We ran all experiments on a 64-processor machine: Intel(R) Xeon(R) CPU E5-4610 v2 @ 2.30GHz,
with 128GB of memory. Multi-threaded experiment code is implemented in Java and executed under
Ubuntu 14.04 LTS. We used MongoDB\footnote{https://www.mongodb.org/} to store and query the 
datasets. Note that all the features are precomputed and saved to this database. This saves us a
tremendous amount of execution time, since feature extraction becomes very time-, memory- and 
storage-consuming, especially, for dynamic features such as Words and Part-of-Speech Tags.
We plan to make all of the source code publicly available prior to publication of this paper.

Run-time complexity of the MLLF algorithm (to link a single unknown account) is 
$O(|A_{known}|*CS_d*|F|)$, which is proportional to:
\begin{compactitem}
  \item Size of the known accounts set
  \item Time to calculate Chi-Square distance between two feature sets
  \item Number of feature categories
\end{compactitem}
Figure \ref{fig:execution} shows two plots, \ref{fig:execution-easy-medium} for
$|A_{known}|$, from $100$ to $10,000$; and \ref{fig:execution-hard} for $|A_{known}|$ from $10,000$ to 
$100,000$. We split plots into two parts since \ref{fig:execution-hard} uses one less feature (11 total), 
as mentioned in Section \ref{subsec:hard}. Also, we are only reporting execution times of 
Flickr$\rightarrow$Twitter and Yelp$\rightarrow$Twitter, since only Twitter has up to $100,000$ authors
in our dataset.

We observe linear trend in both plots, as expected from the algorithm complexity.
Execution time reaches almost $1$ second for $10,000$ authors, and approximately 
$13$ seconds for $90,000$ authors. We observe an exponential jump for $100,000$ authors
This occurs because of insufficient RAM, which forces the code to resort to using the disk swap partition.

\begin{figure}[h]
  \centering
  \includegraphics[width=0.47\textwidth,height=6.5cm]{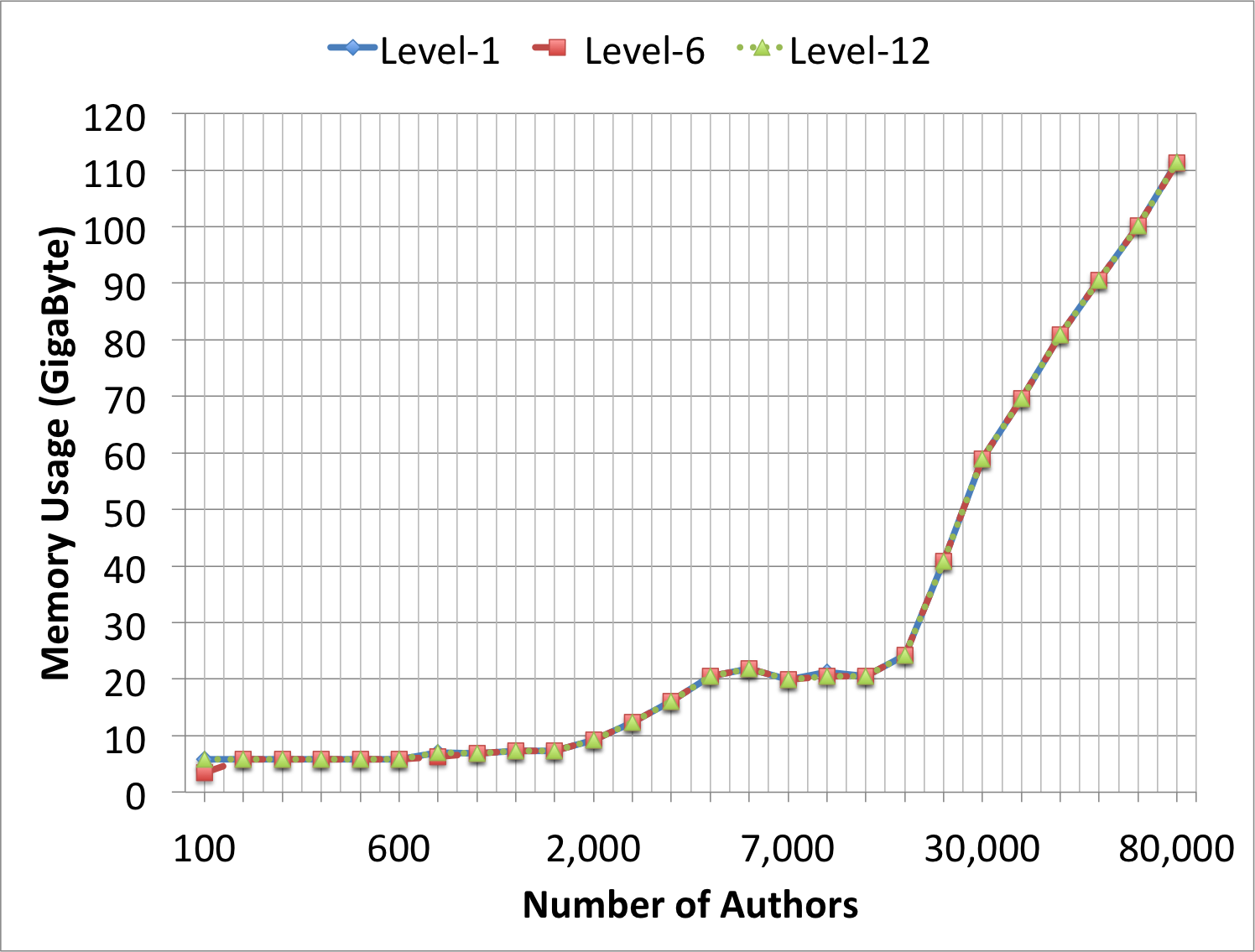}
  \caption{Memory footprint of MLLF running for Flickr$\rightarrow$Twitter (memory consumptions is similar in other OSN combinations) when number of authors increases from $100$ to $80,000$. Each curve refers to a different level in MLLF.}
  \label{fig:memory}
\end{figure}

After the execution times, we present the memory footprint of MLLF in Figure \ref{fig:memory}.
Since running MLLF with more than $90,000$ authors causes disk swap partition usage,
we are only showing memory consumption up to $80,000$ authors.
As expectedly, memory usage increases linearly while author set size grows.
MLLF requires $7$ gigabyte of memory for $1,000$ authors, $24$ gigabyte for $10,000$
authors and $111$ gigabyte for $80,000$ authors. Most important memory characteristics
of MLLF is even though algorithms works in hierarchical increments, memory usage
does not increase after each level. This is because MLLF is using only one feature
category in each level. Thus, conventional algorithms, that uses more than one feature category,
would require a lot more memory than MLLF.

Of course, better software engineering practices would likely lower the memory footprint
and improve execution time. However, we believe that current results give a
general idea of MLLF's scalability. For example, in only $13$ seconds, MLLF
can link an unknown account with 71\% accuracy, within a set of $90,000$ authors.

\subsection{Summary}
Our experimental results can be summarized as follows:
\begin{compactenum}
\item We begin with a baseline method using a greedy hill-climbing algorithm on features to improve linkability. 
This results in 11\% Top-$1$ LR from Twitter$\rightarrow$Yelp, which is comparable to prior results in 
\cite{goga2013exploiting}. We concluded that recent stylometric linkability models are not resilient 
when used to link accounts across heterogeneous OSNs; see Section \ref{subsec:baseline}.
\item We then proposed a new Multi-Level Linker Framework (MMLF), which improves LRs by 
around 50\%; see Section \ref{subsec:framework}.
\item Next, we demonstrated MLLF's scalability when the number of authors grows from $100$ to $100,000$.
We managed to reach Top-$10$ LRs of 68\% for Flickr$\rightarrow$Twitter and 42\% for Yelp$\rightarrow$Twitter 
in a set of $100,000$ possible authors; see Section \ref{subsec:scalability}.
\item Finally, we discussed the run-times and memory requirements of MLLF as the number of authors increases.
MLLF only takes around $8$ seconds to link an unknown account from 
either Flickr or Yelp to Twitter in a set of $80,000$ possible authors,
and requires around $111$ gigabyte of memory; see Section \ref{subsec:execution-time}.
\end{compactenum}
In the next section, we discuss the results in more detail.

\section{Discussion: Hypothetical Q\&A}
\label{sec:discussion}
We now attempt to elaborate (in a Q\&A style) on some potential 
issues prompted by the results described in 
the previous section.

\textbf{How to leverage relative disparity of results?}
Our trilateral account linkability study using simple stylometric features 
concludes that linkability from Flickr or Yelp to Twitter is the highest. 
Meanwhile, linkability from Yelp or Twitter to Flickr is the lowest.
This means that Twitter is the best and Flickr is the worst OSN, respectively, 
as a basis for constructing stylometric profile of users.

\textbf{Why is linkability to Twitter so high?}
Our initial and somewhat intuitive expectation was that linkability to Yelp would be 
the highest, since Yelp, unlike Twitter, does not have text size limits. We 
anticipated that a typical Yelp user exhibits a writing style very similar to that used
in their everyday writing activities. In contrast, Twitter forces certain verbal contortions 
and compressions due to its 140-characters limitation. However, it turns out that 
Twitter allows us to build a better stylometric profile than Yelp. One potential explanation
is restricted context or focus: Twitter is a general-purpose micro-blogging OSN, 
while Yelp is primarily about reviewing restaurants, hotels and various other venues.
In Twitter, people write mostly about themselves, other people, 
events (e.g., news), yet the context is totally unrestricted, i.e., anything goes. 
This could mean that contextual freedom allows capturing one's writing style better 
as long as a user authors a sufficient overall amount of text.

\textbf{Why is Flickr$\rightarrow$Twitter linkability higher than\\ 
Yelp$\rightarrow$Twitter?}
One possible reason is that, like tweets, photo annotations tend to be relatively short, albeit
without a mandatory upper limit on words or characters. In contrast, free-form reviews can 
(and often are) quite long. Therefore, Flickr encourages a certain writing style that is somehow 
closer to Twitter than that of Yelp.

\textbf{Why is linkability to Flickr the lowest? Can it be improved?}
Our conclusion regarding Flickr is that photo descriptions are simply not rich 
enough to build one's accurate stylometric profile. We suppose that people mostly 
write general facts about photos they share, and do not provide really personalized text.
How to better utilize photo annotations to improve linkability remains an interesting open 
question. Perhaps we need a significantly larger body of text along with photo annotations.
One possible approach would be to crawl Instagram profiles of our matching accounts and 
combine them with Flickr profiles. This would yield a larger body of text, which could increase 
LRs. As a separate item, studying linkability between Flickr and Instagram -- two OSNs similar
in their mission -- will be another interesting future work direction.

\textbf{Can MLLF scale to millions of accounts?}
MLLF's complexity increases linearly with the number of accounts. Therefore,
we believe it can be used in a much bigger account set, given enough RAM.
According to the trend observed in our experiment execution times, we estimate that
it would take around $2.5$ minutes to link one unknown account to $1,000,000$ 
known ones. Of course memory footprint, multi-threading and implementation efficiency
can be further optimized using better software engineering practices, which we also 
leave to future work.

\textbf{How should features be ordered in a real world study?}
Current implementation of MLLF shuffles available features and uses a different
feature in each level. One can imagine that if a feature is weak and is unfortunately
chosen in early levels, then the true match will be filtered out.
As part of our future work, we plan to provide heuristics to order features
so that linkability will be maximized.
But right now, our suggestion is to order features randomly and run MLLF multiple times.
We averaged our linkability results with 10 random ordering of features and
linkability is already highly accurate.

\textbf{Can we use other stylometric features?}
Extending MLLF's feature set with other Writeprints features is very likely to influence
LRs. As part of future work we plan to gradually experiment with the other 12 Writeprints 
features.

\textbf{Can other textual OSN features be used?}
Hashtags in Twitter and tags in Flickr are examples of textual OSN features that we
excluded in this study. They provide a mechanism for labeling each post, which is useful for 
classifying and finding interests. Also, they are generally not authored by the person who 
uses them in tweets or annotations, respectively. Therefore, they cannot be directly considered 
as part of a user's stylometric profile. However, a recent Twitter-based study \cite{mish-wpes-14} 
demonstrated a technique which combines hashtags with other stylometric features to improve
linkability. We believe that a similar approach might also be helpful in our settings.
However, we note that not all OSNs support labeling, e.g., Yelp does not.

\textbf{Can MLLF be combined with other classifiers?}
We would like to extend MLLF with other types of techniques, such as 
SVM, Na\"ive Bayes and $k$-nearest neighbors. The intuition is that 
these more complex and expensive methods can be plugged in at the highest 
level of MLLF, where we currently have the lowest number of known accounts.
This might keep execution overhead of a more complex method minimal, and 
increase LRs.

\textbf{Can two OSN profiles be combined while linking to an unknown account?}
We do not yet know how combining homogeneous and/or heterogeneous accounts influences
linkability. This is another open question. One obvious step is to combine Yelp and Twitter 
profiles of known accounts, while trying to link to an unknown Flickr account.
Such a hypothetical system could generate a generic stylometric fingerprint, which would be 
a real breakthrough in author attribution and linkability.

\textbf{What can be said about linkability in the context of a generic OSNs?}
We believe our trilateral linkability study is only the first step. It is natural to add other (including 
different types of) OSNs, in particular, a global general-purpose OSN, such as 
Facebook, Google+, or LinkedIn. Once again, this is an item for near-term future work.

\section{Conclusions}
\label{sec:conclusion}
Despite the elusiveness of OSN privacy, many users expect that multiple 
accounts they operate within one, and on more than one, OSNs remain 
isolated, i.e., unlinkable, owing perhaps to very different OSN missions. 
For example, photo-sharing, micro-blogging and product/service reviews 
appear to be quite distinct types of OSN specialization.
However, this is unfortunately not the case, as supported by the results 
of the study presented in this paper. It also represents the first large-scale stylometric-based account
linkability experiment conducted across three heterogeneous OSNs: 
Yelp, Twitter and Flickr. 
\section*{ACKNOWLEDGMENTS}
We are very grateful to the authors of \cite{goga2013exploiting} for kindly
sharing with us the crawled Yelp, Flickr, and Twitter dataset used in their
previous work.

\bibliographystyle{acm}
\bibliography{paper} 

\begin{thebibliography}{10}

\bibitem{aol}
{AOL}.
\newblock {http://www.aol.com. Last accessed on 2015-04-30}.

\bibitem{flickr}
{Flickr}.
\newblock {https://www.flickr.com. Last accessed on 2015-04-27}.

\bibitem{twitter}
{Twitter}.
\newblock {https://www.twitter.com. Last accessed on 2015-04-26}.

\bibitem{twitterusers}
{Twitter Blog}.
\newblock {https://blog.twitter.com/2013/celebrating-twitter7. Last accessed on
  2015-04-26}.

\bibitem{yelp}
{Yelp}.
\newblock {http://www.yelp.com. Last accessed on 2015-04-23}.

\bibitem{yelpsabout}
{Yelp -- About Us}.
\newblock {http://www.yelp.com/about}.

\bibitem{yelpeliete}
{Yelp Elite Squad}.
\newblock {http://www.yelp.com/elite. Last accessed on 2015-04-23}.

\bibitem{abbasi2008writeprints}
{\sc Abbasi, A., and Chen, H.}
\newblock Writeprints: A stylometric approach to identity-level identification
  and similarity detection in cyberspace.
\newblock {\em ACM Transactions on Information Systems (TOIS) 26}, 2 (2008), 7.

\bibitem{afroz2014doppelganger}
{\sc Afroz, S., Islam, A.~C., Stolerman, A., Greenstadt, R., and McCoy, D.}
\newblock Doppelg{\"a}nger finder: Taking stylometry to the underground.
\newblock In {\em Security and Privacy (SP), 2014 IEEE Symposium on\/} (2014),
  IEEE, pp.~212--226.

\bibitem{mish-wpes-14}
{\sc Almishari, M., Kaafar, D., Oguz, E., and Tsudik, G.}
\newblock {Stylometric Linkability of Tweets}.
\newblock In {\em WPES\/} (2014).

\bibitem{almishari2014fighting}
{\sc Almishari, M., Oguz, E., and Tsudik, G.}
\newblock Fighting authorship linkability with crowdsourcing.
\newblock In {\em Proceedings of the second edition of the ACM conference on
  Online social networks\/} (2014), ACM, pp.~69--82.

\bibitem{derczynski2013twitter}
{\sc Derczynski, L., Ritter, A., Clark, S., and Bontcheva, K.}
\newblock Twitter part-of-speech tagging for all: Overcoming sparse and noisy
  data.
\newblock In {\em RANLP\/} (2013), pp.~198--206.

\bibitem{you-what-u-say}
{\sc Frankowski, D., Cosley, D., Sen, S., Terveen, L., and Riedl, J.}
\newblock {You Are What You Say: Privacy Risks of Public Mentions}.
\newblock In {\em International ACM SIGIR Conference on Research and
  Development in Information Retrieval\/} (2006).

\bibitem{goga2013exploiting}
{\sc Goga, O., Lei, H., Parthasarathi, S. H.~K., Friedland, G., Sommer, R., and
  Teixeira, R.}
\newblock Exploiting innocuous activity for correlating users across sites.
\newblock In {\em Proceedings of the 22nd international conference on World
  Wide Web\/} (2013), International World Wide Web Conferences Steering
  Committee, pp.~447--458.

\bibitem{iofciu2011identifying}
{\sc Iofciu, T., Fankhauser, P., Abel, F., and Bischoff, K.}
\newblock Identifying users across social tagging systems.
\newblock In {\em ICWSM\/} (2011).

\bibitem{footprint}
{\sc Irani, D., Webb, S., Li, K., and Pu, C.}
\newblock Large online social footprints--an emerging threat.
\newblock In {\em CSE '09: Proceedings of the 2009 International Conference on
  Computational Science and Engineering\/} (Washington, DC, USA, 2009), IEEE
  Computer Society, pp.~271--276.

\bibitem{kevselj2003n}
{\sc Ke{\v{s}}elj, V., Peng, F., Cercone, N., and Thomas, C.}
\newblock N-gram-based author profiles for authorship attribution.
\newblock In {\em Proceedings of the conference pacific association for
  computational linguistics, PACLING\/} (2003), vol.~3, pp.~255--264.

\bibitem{koppel2005automatically}
{\sc Koppel, M., Schler, J., and Zigdon, K.}
\newblock Automatically determining an anonymous author's native language.
\newblock In {\em Intelligence and Security Informatics}. Springer, 2005,
  pp.~209--217.

\bibitem{mcdonald2012use}
{\sc McDonald, A.~W., Afroz, S., Caliskan, A., Stolerman, A., and Greenstadt,
  R.}
\newblock Use fewer instances of the letter ``i": Toward writing style
  anonymization.
\newblock In {\em Privacy Enhancing Technologies\/} (2012), Springer,
  pp.~299--318.

\bibitem{mish-linkability}
{\sc Mishari, M.~A., and Tsudik, G.}
\newblock Exploring linkability of user reviews.
\newblock In {\em ESORICS\/} (2012).

\bibitem{herbert-deanonymizer}
{\sc Nanavati, M., Taylor, N., Aiello, W., and Warfield, A.}
\newblock {Herbert West -- Deanonymizer}.
\newblock In {\em 6th USENIX Workshop on Hot Topics in Security\/} (2011).

\bibitem{internet-scale}
{\sc Narayanan, A., Paskov, H., Gong, N.~Z., Bethencourt, J., Stefanov, E.,
  Shin, E. C.~R., and Song, D.}
\newblock {On the Feasibility of Internet-Scale Author Identification}.
\newblock In {\em IEEE Symposium on Security and Privacy\/} (2012).

\bibitem{deanonymize-netflix}
{\sc Narayanan, A., and Shmatikov, V.}
\newblock {Robust De-anonymization of Large Sparse Datasets}.
\newblock In {\em IEEE Symposium on Security and Privacy\/} (2009).

\bibitem{dperitoPETS11}
{\sc Perito, D., Castelluccia, C., Kaafar, M.~A., and Manils, P.}
\newblock {How Unique and Traceable Are Usernames?}
\newblock In {\em PETS\/} (2011).

\bibitem{rao2000can}
{\sc Rao, J.~R., and Rohatgi, P.}
\newblock Can pseudonymity really guarantee privacy?
\newblock In {\em USENIX Security Symposium\/} (2000).

\bibitem{author-survey}
{\sc Stamatatos, E.}
\newblock {A Survey of Modern Authorship Attribution Methods}.
\newblock In {\em Journal of the American Society for Information Science and
  Technology\/} (2009).

\bibitem{toutanova2003feature}
{\sc Toutanova, K., Klein, D., Manning, C.~D., and Singer, Y.}
\newblock Feature-rich part-of-speech tagging with a cyclic dependency network.
\newblock In {\em Proceedings of the 2003 Conference of the North American
  Chapter of the Association for Computational Linguistics on Human Language
  Technology-Volume 1\/} (2003), Association for Computational Linguistics,
  pp.~173--180.

\end{thebibliography}
{\renewcommand{\baselinestretch}{1}
\onecolumn

\section*{APPENDIX A: Linkability Algorithm Pseudo-Code}
\label{sec:appendix-framework}
\vspace*{0.5cm}

\begin{algorithm*}[!htp] 
\small
  \begin{algorithmic}[1] 
    \Require $DB$; Database interface to access dataset
    \Require $OSN_U$; OSN type of the unknown accounts
    \Require $OSN_K$; OSN type of the known accounts
    \Require $authorSize$; Number of authors in $OSN_K$
    \State // Get all the available features and shuffle them.
    \State $features\gets shuffle(getAllFeatures())$
    \\
    \State // Get all the matching accounts in between $OSN_U$ and $OSN_K$.
    \State // We use all the matching accounts as shown from Table \ref{table:matching_accounts}.
    \State $matchingAccounts\gets DB.getMatchingAccounts(OSN_U, OSN_K)$
    \\
    \State // Get $authorSize$ number of known accounts from $OSN_K$.
    \State $possibleAccounts\gets DB.getPossibleAccounts(authorSize, OSN_K)$
    \\
    \State // Initialize other variables.
    \State $level\gets 0$
    \Comment{Initialize experiment level to beginning}
    \State $results\gets$ Map<Account, List<Result$\textgreater\textgreater$
    \Comment{$results$ is a map of Account to a list of experiment Result objects}
    \\
    \State // Do the level-0 experiment with features[0].
    \State $experiment\gets Experiment(matchingAccounts, possibleAccounts)$
    \State $result\gets experiment.run(features[0])$
    \Comment{Perform the level-0 experiment}
    \\
    \State // For each matchingAccount, create a list from her experiment result and save it to $results$.
    \For{$(matchingAccount$ in $matchingAccounts)$}
      \State $results[matchingAccount]\gets $List<Result>($result$.get($matchingAccount$))
    \EndFor
    \\
    \State $topT\gets authorSize$
    \Comment{$topT$ is used to filter out accounts in each experiment level}
    \State $level\gets level+1$
    \Comment{Increase the $level$ since we are done with level-0 experiment}
    \\
    \State // Continue with next level of experiments using remaining features.
    \For{($level < features.length$)}
      \\
      \State $feature\gets features[level]$
      \Comment{Get the new feature to be used in this experiment $level$}
      \State $topT\gets TopT/2$
      \Comment{In each level, filter out the halve of possible accounts}
      \State $experiments\gets List$<Experiment>
      \Comment{$experiments$ is a list of Experiment objects}
      \\
      \State // For every $matchingAccount$, assess the latest linkability result and try to improve it using this $feature$ of this level.
      \For{$(matchingAccount$ in $matchingAccounts)$}
        \\
        \State // If the latest level of result for current $matchingAccount$ reported $topT$ linkability,
        \State // then try to improve this result using $feature$.
        \If{($results[matchingAccount].getTopElement().getPosition() < topT$)}
          \\
          \State // Create a new experiment where $matchingAccount$ is the only unknown account
          \State // and $topT$ possible authors for $matchingAccount$ are the possible authors.
          \State $experiment\gets $Experiment$(matchingAccount, matchingAccount.getTopTAccounts(topT))$
          \\
          \State // Run a new experiment with a new feature
          \State $result\gets experiment$.run($feature$)
          \\
          \State // Append the latest experiment result to previous list of results of $matchingAccount$.
          \State $results[matchingAccount]$.append($result$)
        \Else
          \State // There are no possible improvements, linkability of this $matchingAccount$ will be reported as it is.
        \EndIf
      
      \EndFor
      \State $level\gets level+1$
      \Comment{Increase the experiment level}
    \EndFor

    \State \textbf{return} $results$ \Comment{Return the map of experiment results}
  \end{algorithmic}
  \caption{Experiment algorithm, linkability from $OSN_U$ to $OSN_K$}
  \label{algo:linker}
\end{algorithm*}}

\end{document}